\documentclass{article}
\usepackage{amssymb}
\usepackage{amsfonts}
\usepackage{amsmath}
\usepackage{hyperref}
\usepackage{graphicx}

\hypersetup{
colorlinks=true,
citecolor=blue,
linkcolor=blue,
urlcolor=blue
}

\begin{document}

\title{\textbf{Five-dimensional Einstein-Chern-Simons cosmology}}
\author{ F. G\'{o}mez$^{1}$\thanks{
fernando.gomez@ulagos.cl }\ , S. Lepe$^{2}$\thanks{%
samuel.lepe@pucv.cl }\ , C. Quinzacara$^{3}$\thanks{%
cristian.cortesq@uss.cl }\ \ and P. Salgado$^{4}$\thanks{%
patsalgado@unap.cl} \\
$^{1}$Departamento de Ciencias Exactas, Universidad de Los Lagos, \\
Avenida Fuchslocher 1305, Osorno, Chile\\
$^{2}$Instituto de F\'{\i}sica, Pontificia Universidad Cat\'{o}lica de
Valpara\'{\i}so,\\
Avenida Brasil 2950, Valpara\'{\i}so, Chile. \\
$^{3}$Facultad de Ingenier\'{\i}a y Tecnolog\'{\i}a, Universidad San Sebasti%
\'{a}n,\\
Lientur 1457, Concepci\'{o}n 4080871, Chile\\
$^{4}$Instituto de Ciencias Exactas y Naturales, Facultad de Ciencias\\
Universidad Arturo Prat, Avenida Arturo Prat 2120, Iquique, Chile }
\maketitle

\begin{abstract}
We consider a five-dimensional Eins\-tein-Chern-Si\-mons action which is composed of a gravitational sector and a sector of matter, where the gravitational sector is given by a Chern-Simons gravity action instead of the Einstein-Hilbert action, and where the matter sector is given by a perfect fluid. The gravitational lagrangian is obtained gauging some Lie-algebras, which in turn, were obtained by S-expansion procedure of Anti-de Sitter and de Sitter algebras. On the cosmological plane, we discuss the field equations resulting from the Anti-de Sitter and de Sitter frameworks and we show analogies with four-dimensional cosmological schemes.
\end{abstract}

\section{Introduction}

The construction of a gauge theory of gravity requires an action that does not consider a fixed space-time background. An Lagrangian for gravity
fulfilling these conditions is given by a Chern-Simons (ChS) form for the (Anti-)de Sitter algebra. The corresponding Chern-Simons action can be
written as
\begin{equation}
S_{(\mathrm{A})\mathrm{dS} }^{\left( 5\right) }=\frac{1}{8\kappa}\int \pm \frac{1}{5l^{2}}\varepsilon_{abcde}e^{a}e^{b}e^{c}e^{d}e^{e}+\frac{2}{3}\varepsilon_{abcde}R^{ab}e^{c}e^{d}e^{e}\pm l^{2}\varepsilon_{abcde} R^{ab}R^{cd}e^{e} ,  \label{1}
\end{equation}
which is off-shell invariant under the AdS-Lie algebra $\mathfrak{so}\left(4,2\right)$ and dS-Lie algebra $\mathfrak{so}\left(5,1\right)$ respectively. Here, $e^{a}$ corresponds to the 1-form \emph{vielbein}, and $R^{ab}=\text{d}\omega ^{ab}+\omega _{\ c}^{a}\omega ^{cb}$ to the curvature 2-form in the first-order formalism obtained from 1-form spin connection $\omega ^{ab}$, which in turn is related to the Riemann curvature tensor.

From lagrangian (\ref{1}) it is apparent that neither the $l\rightarrow\infty $ nor $l\rightarrow 0$ limit yield the Einstein-Hilbert term $\varepsilon _{a_{1}\cdots a_{5}}R^{a_{1}a_{2}}e^{a_{3}}\cdots e^{a_{5}}$ alone. Such limits will lead either to the Gauss-Bonnet term or to the cosmological constant term by itself, respectively.

If ChS theories are the appropriate gauge-theories to provide a framework for the gravitational interaction, then these theories must satisfy the
correspondence principle, namely they must be related to General Relativity (GR). 

Some time ago was shown that the standard, 5-dimensional (5D) GR can be obtained from ChS gravity theory for a \textit{certain} Lie algebra $\mathfrak{B}_{5}$, whose generators satisfy the commutation relation shown in equation (7) of Ref. \cite{gomez}. This algebra was obtained from the Anti-de Sitter (AdS) algebra and a particular semigroup $S$ by means of the S-expansion procedure introduced in Refs. \cite{salg2}, \cite{salg3}, \cite{azcarr}.

It is straightforward to prove that GR in 5D can be also obtained from a ChS gravity for an Lie algebra $\mathfrak{\tilde{B}}_{5}$, which is obtained
from the de Sitter (dS) algebra using the mentioned S-expansion method. The only difference between $\mathfrak{B}_{5}$ and $\mathfrak{\tilde{B}}_{5}$ is
found in the sign of commutator between generalized translations $\left[\boldsymbol{P}_{a} , \boldsymbol{P}_{b} \right]$ from the algebra shown in equation (7) in Ref. \cite{gomez}.

Using theorem~VII.2 of Ref.~\cite{salg2} and using the extended Cartan's homotopy formula as in Ref. \cite{salg4}, was found that the ChS Lagrangian
in five dimensions for the $\mathfrak{B}_{5}$ and $\mathfrak{\tilde{B}}_{5}$ algebras is given by 
\begin{align}
L_{\text{EChS}}^{(5)}& =\alpha _{1}\left( \pm l^{2}\right) \varepsilon
_{abcde}e^{a}R^{bc}R^{de}  \notag \\
& \quad +\alpha _{3}\varepsilon _{abcde}\left( \frac{2}{3}%
R^{ab}e^{c}e^{d}e^{e}+\left( \pm l^{2}\right) R^{ab}R^{cd}h^{e}+2\left( \pm
l^{2}\right) k^{ab}R^{cd}T^{e}\right) ,  \label{2}
\end{align}
where $\alpha _{1}$, $\alpha _{3}$ are parameters of the theory, $l$ is a coupling constant, $R^{ab}$ is the 2-form curvature which we have already introduced, $e^{a}$ is
the \textit{vielbein} and, $h^{a}$ and $k^{ab}$ are others gauge fields presents in the theory \cite{salg1}.

From here we can see that the limit where $l\longrightarrow 0$ leads us to just the Einstein-Hilbert dynamics in the vacuum (for details see \cite{salg1}). The $h^{a}$ field has analogous properties to \textit{vielbein} $e^{a}$. Both of them are vectors under local Lorentz transformations generated by $\boldsymbol{J}_{ab}$ and represent spin-2 bosons in the first order formalism. In the case of vielbein, it is graviton and the case of $h^{a}$, it is undetermined field. This \textquotedblleft duality" is retained at lagrangian level (\ref{2}) and at its functional variation. In fact, we note both fields couple to a term which is quadratic in the curvature $R^{ab}$. A completely analogous situation is also found between spin
connection $\omega ^{ab}$ and $k^{ab}$ field.

To determine the constants $\alpha _{1}$ and $\alpha
_{3} $ let's study the equation the equation (\ref{2}) in some detail:

(a) Carrying out a generalized In\"{o}n\"{u}-Wigner contraction (the inverse procedure to the S-expansion process) of the algebras  $\mathfrak{\tilde{B}}_{5}$ and $\mathfrak{B}_{5}$ we obtain the algebra (A)dS. Likewise, the application of the contraction method to the lagrangian (\ref{2}) we obtain the lagrangian corresponding to the action (\ref{1}). This means that in this limit $\alpha _{1}l^{2}$ must match $l^{2}/8\kappa $, i.e., $\alpha _{1}=1/8\,\kappa$.

(b) On the other hand the Einstein-ChS Lagrangian $L_{\text{EChS}}^{(5)}$ has the interesting property that leads to the Einstein-Hilbert lagrangian terms of the 5-dimensional ChS-AdS lagrangian (\ref{1}) in the limit where the coupling constant $l$ tends to zero. This means that in this limit $\left(2/3\right) \alpha _{3}$ must match $2/24\,\kappa$, i.e., $\alpha
_{3}=1/8\,\kappa$.

In the present work we study the cosmological consequences of the field equations corresponding to the Lagrange function (\ref{2}), where the $``+"$
sign corresponds to the Lagrangian obtained by gauging the algebra $\mathfrak{B}_{5}$, which is obtained by expansion of the AdS-algebra and
the sign $``-"$ corresponds to the Lagrangian obtained by gauging the algebra $\mathfrak{\tilde{B}}_{5}$, which is obtained by expansion of the dS-algebra.

We will solve the field equations for a (spatially) flat Friedmann-Lem\^{a}itre-Robertson-Walker (FLRW) metric. In the field equations corresponding to the signs $``\pm"$, a term proportional to $3l^{2}H^{4}$ appears ($H$ is the Hubble parameter) coming from the quadratic curvature corrections introduced by lagrangian (\ref{2}). This term will be discussed in both schema ($``+"$ and $``-"$) in order to describe the resultant cosmologies.

The article is organized as follows:  in Section II we consider a total action composed by the EChS action (\ref{2}), with $\alpha _{1}=\alpha _{3}=1/8\kappa$, and an action for matter, and
then we derive the corresponding field equations. Under certain impositions on the fields, we obtain the equations for the 5-dimensional FLRW spacetime, and we study some general cosmological consequences of this schema in the flat case. In this framework, in Section III, we find the Hubble parameter and the deceleration parameter in several cases, for both $\mathfrak{\tilde{B}}_{5}$ and $\mathfrak{B}_{5}$ EChS cosmologies, and we compare these results with their 4D analogues. Finally, concluding remarks are presented in Section $IV$.

\section{Five-dimensional EChS-FRLW field equa\-tions}
We start our study by considering the composed total action 
\begin{equation}
S=S_{\text{EChS}}^{(5)}+S_{M},  \label{6}
\end{equation}
where $S_{\text{EChS}}^{(5)}$ is the action formed from lagrangian (\ref{2}) and $S_{M}=S_{M}(e^{a}, h^{a}, \omega ^{ab},\break k^{ab}, \ldots)$ is an action for matter. The variation of the action (\ref{6}) leads to the following field equations
\begin{align}
\varepsilon _{abcde}\left( 2R^{ab}e^{c}e^{d}\pm l^{2}R^{ab}R^{cd}\pm 2l^{2}
\mathrm{D}_{\omega }k^{ab}R^{cd}\right) &=-8\kappa \frac{\delta L_{M}}{%
\delta e^{e}},  \label{7} \\
\pm l^{2}\varepsilon _{abcde}R^{ab}R^{cd} &=-8\kappa \frac{\delta L_{M}}{%
\delta h^{e}},  \label{8} \\
\pm\varepsilon _{abcde}\left( l^{2}R^{cd}T^{e}+ l^{2}\mathrm{D} k^{cd}T^{e} \pm e^{c}e^{d}T^{e}+ l^{2}R^{cd}\mathrm{D}h^{e}+ l^{2}R^{cd}k_{\ f}^{e}e^{f}\right) &=-4\kappa \frac{\delta L_{M}}{\delta \omega
^{ab}},  \label{9} \\
\pm l^{2}\varepsilon _{abcde}R^{cd}T^{e} &=-4\kappa \frac{\delta L_{M}}{%
\delta k^{ab}}, \label{10}
\end{align}
where $L_M$ is the corresponding matter lagrangian.

For simplicity, we assume the torsionless condition 
\begin{equation}
T^{a}=\mathrm{d}e^{a}+\omega_{\ b}^{a}e^{b}=0,  \label{11}
\end{equation}
and a null $k^{ab}$-field. However, due to the gauge freedom, the last condition ($k^{ab} = 0$) on field equations is not too restrictive as we could think at first sight. On the other hand, the first condition ($T^a = 0$) leads to $\delta L_M/\delta k^{ab}=0$. So, both conditions are consistent with each other. Also, we assume $\delta L_{M}/\delta \omega ^{ab}=0$ which is a consequence of considering spinless matter 
\begin{equation}
\frac{\delta L_{M}}{\delta \omega ^{ab}}=\frac{1}{4!}\mathcal{S}_{ab}^{c}\,\varepsilon _{cdefg}\,e^{d}e^{e}e^{f}e^{g},
\end{equation}
where $\mathcal{S}_{\ ab}^{c}$ is the spin tensor. In this case, the above equations take the form 
\begin{eqnarray}
\varepsilon _{abcde}\left( 2R^{ab}e^{c}e^{d}\pm l^{2}R^{ab}R^{cd}\right)
&=&-8\kappa \frac{\delta L_{M}}{\delta e^{e}},  \label{12} \\
\pm l^{2}\varepsilon _{abcde}R^{ab}R^{cd} &=&-8\kappa \frac{\delta L_{M}}{%
\delta h^{e}},  \label{13} \\
\pm l^{2}\varepsilon _{abcde}R^{cd}\mathrm{D}h^{e} &=&0,  \label{14}
\end{eqnarray}%
where $\kappa $ is a coupling constant related to the effective Newton constant.

The quantities $\delta L_{M}/\delta e^{a}$ and $\delta L_{M}/\delta h^{a}$ can be expressed as
\begin{equation}
\frac{\delta L_{M}}{\delta e^{a}}=\frac{1}{4!}T_{\ a}^{b}\varepsilon
_{bcdef}\,e^{c}e^{d}e^{e}e^{f}\text{ \ \ \ },\text{ \ \ \ }\frac{\delta L_{M}%
}{\delta h^{a}}=-\frac{1}{4!}T_{\ \ \ \ \ a}^{(h)b}\varepsilon
_{bcdef}\,e^{c}e^{d}e^{e}e^{f},  \label{15}
\end{equation}%
where $T_{\ a}^{b}$ represents a usual energy-momentum tensor and $T_{\ \ \
\ \ a}^{(h)b}$ represents its counterpart for the $h$-field (note that, as we will see later, the introduction of $T_{\ \ \ \ \ a}^{(h)b}$ as a kind of
energy-momentum tensor is not entirely accurate, but it is convenient to shed light on the nature of the hitherto unknown $h$-field). Therefore, the system (\ref{12})-(\ref{14}) takes the form 
\begin{align}
\varepsilon _{abcde}\left( 2R^{bc}e^{d}e^{e}\pm l^{2}R^{bc}R^{de}\right) & =-%
\frac{\kappa }{3}T_{\ a}^{b}\varepsilon _{bcdef}\,e^{c}e^{d}e^{e}e^{f},
\label{17} \\
\pm l^{2}\varepsilon _{abcde}R^{bc}R^{de}& =\frac{\kappa }{3}T_{\ \ \ \ \
a}^{(h)b}\varepsilon _{bcdef}\,e^{c}e^{d}e^{e}e^{f},  \label{18} \\
\varepsilon _{abcde}R^{cd}\mathrm{D}_{\omega }h^{e}& =0.  \label{19}
\end{align}%
Note that the quantity $T_{a}^{(h)\text{ }b}$ is
conserved, in the same manner as the energy-momentum tensor is conserved; i.e., the equation $\nabla _{\mu }T_{\nu }^{\left( h\right) \mu}=0$ is satisfied.

In order to find cosmological solutions for this system, we will shape the fields involved according to the spatial homogeneity and isotropy suggested
by the cosmological principle. Thus, we consider the 5D-FLRW metric 
\begin{equation}
ds^{2}=-\,\mathrm dt^{2}+a^{2}(t)\left[ \frac{\mathrm  dr^{2}}{1-kr^{2}}+r^{2}\left(\mathrm d\theta
_{1}^{2}+\sin^{2}\theta _{1}\,\mathrm d\theta _{2}^{2}+\sin^{2}\theta _{1}\sin^{2}\theta
_{2}\,\mathrm d\theta _{3}^{2}\right) \right] ,  \label{19.2}
\end{equation}%
where $a(t)$ is the cosmic scale factor, $k=+1,0,-1$ and we choose the vielbein as
\begin{align}
e^{0}=\mathrm dt\text{ \ \ },\text{ \ }e^{1}=\frac{a(t)}{\sqrt{1-kr^{2}}}\,\mathrm dr\text{ \ \ \ } , \text{ \ }e^{2}=a(t)r\,\mathrm d\theta _{1},\notag\\
e^{3}=a(t)r\sin\theta _{1}\,\mathrm d\theta _{2}\text{ \ \ \ },\text{ \ \ }%
e^{4}=a(t)r\sin\theta _{1}\sin\theta _{2}\,\mathrm d\theta _{3}.  \label{19.3}
\end{align}
On the other hand, the $h$-field is expressed as (see
Appendix)
\begin{equation}
\begin{array}{c}
h^{0}=C_{0}e^{0}, \\ 
h^{p}=h(t)e^{p}\text{ \ \ },\text{ \ \ }p=1,2,3,4.%
\end{array}
\label{19.4}
\end{equation}%
where $C_{0}$ is a constant and $h(t)$ is a time-dependent function. In addition, we assume that the material content of the model is described by perfect fluids
\begin{equation}
T_{\mu }^{\text{ \ }\nu }=\mathrm{diag}(-\rho ,p,p,p,p),  \label{33}
\end{equation}
\begin{equation}
T_{\mu }^{(h)\nu }=\mathrm{diag}(-\rho ^{(h)},p^{(h)},p^{(h)},p^{(h)},p^{(h)}),
\label{34}
\end{equation}%
being $\rho $ and $p$ the usual
energy density and pressure and, $\rho^{(h)}$ and $p^{(h)}$ represent their counterparts corresponding
to field the $h^{a}$.

The EChS field equations for the 5D-FLRW metric are obtained from the replacement of Eqs. (\ref{19.3})-(\ref{34}) in Eqs. (\ref{17})-(\ref{19}). In fact, the complete system of equations is given by
\begin{equation}
6\left( H^{2}+\frac{k}{a^{2}}\right) \pm 3l^{2}\left( H^{2}+\frac{k}{a^{2}}
\right) ^{2}=\kappa \rho ,  \label{34.2}
\end{equation}
\begin{equation}
3\left( \dot{H}+2H^{2}+\frac{k}{a^{2}}\right) \pm 3l^{2}\left( \dot{H}%
+H^{2}\right) \left( H^{2}+\frac{k}{a^{2}}\right) =-\kappa p,  \label{34.3}
\end{equation}
\begin{equation}
\rho ^{(h)}=\mp \frac{3l^{2}}{\kappa }\left( H^{2}+\frac{k}{a^{2}}\right)
^{2},  \label{34-4}
\end{equation}
\begin{equation}
p^{(h)}=\pm \frac{3l^{2}}{\kappa }\left( \dot{H}+H^{2}\right) \left( H^{2}+%
\frac{k}{a^{2}}\right),   \label{34-5}
\end{equation}%
\begin{equation}
\left( H^{2}+\frac{k}{a^{2}}\right) \left[ \dot{h}+H\left( h-C_{0}\right) %
\right] =0,  \label{34.6}
\end{equation}%
where $H=\dot{a}/a$ is the Hubble parameter. The
detailed derivation of this system is found in the Appendix.

In this article, we are interested in the flat case. So, if $k=0$, and considering $\kappa =1$, the equations (\ref{34.2}) and (\ref{34.3}) lead, for EChS-$\mathfrak{B}_{5},$ to
\begin{equation}
6H^{2}\left( 1+\frac{1}{2}l^{2}H^{2}\right) =\rho \text{ \ },\text{ \ }3%
\left[ \left( 1+l^{2}H^{2}\right) \dot{H}+\left( 2+l^{2}H^{2}\right) H^{2}
\right] =-p,  \label{35}
\end{equation}%
and for EChS-$\mathfrak{\tilde{B}}_{5},$ to
\begin{eqnarray}
6H^{2}\left( 1-\frac{1}{2}l^{2}H^{2}\right)  &=&\rho \text{ \ },\text{ \ }3%
\left[ \left( 1-l^{2}H^{2}\right) \dot{H}+\left( 2-l^{2}H^{2}\right) H^{2}%
\right] =-p,  \label{36'} \\
\dot{\rho}+4H(\rho +p) &=&0,
\end{eqnarray}%
such that, when $l\rightarrow 0$ we obtain
\begin{equation}
6H^{2}\approx \rho \text{ \ and \ }L_{\text{EChS}}^{(5)}\rightarrow 
\frac{1}{12}\epsilon _{abcde}R^{ab}e^{c}e^{d}e^{e}\text{ },  \label{37}
\end{equation}
and when $l\rightarrow \infty $ we have
\begin{equation}
3l^{2}H^{4}\approx \rho \text{ \ and \ }L_{\text{EChS}%
}^{(5)}\rightarrow \pm \frac{1}{8}l^{2}\epsilon _{abcde}\left(
e^{a}R^{bc}R^{de}+R^{ab}R^{cd}h^{e}\right) .  \label{38}
\end{equation}%
Obviously, in the EChS-$\mathfrak{\tilde{B}}_{5}$ case, the limit $l\rightarrow \infty $ does not make sense. 

If we consider the barotropic equation of state $p=\omega \rho $ and $\omega=\text{const.}$, we obtain
\begin{equation}
\rho \left( z\right) =\rho \left( 0\right) \left( 1+z\right) ^{4\left(
1+\omega \right) },  \label{39}
\end{equation}
where $z=-1+a_{0}/a$ is the redshift parameter. So that when $l\rightarrow 0$ we have
\begin{equation}
H\left( z\right) \rightarrow \left[ \frac{1}{6}\rho \left( 0\right) \right]
^{1/2}\left( 1+z\right) ^{2\left( 1+\omega \right) },  \label{40}
\end{equation}%
and when $l\rightarrow 0$ we find
\begin{equation}
l\rightarrow \infty\ \ \Longrightarrow\ \ H\left( z\right) \rightarrow \left[ 
\frac{1}{3l^{2}}\rho \left( 0\right) \right] ^{1/4}\left( 1+z\right)
^{1+\omega }.  \label{41}
\end{equation}
Now, from the expressions given for $\rho ^{\left( h\right) }$ and $p^{\left( h\right) }$, and considering $k=0$, it is straightforward to show
that
\begin{equation}
p^{\left( h\right) }=q\rho ^{\left( h\right) },
\end{equation}%
where $q$ is the deceleration parameter defined as $q=-1-\dot{H}/H^{2}$. As we will see, the interpretation of $\rho ^{\left( h\right) }$ as a fluid only makes sense in the  EChS-$\mathfrak{\tilde{B}}_{5}$ framework.

On the other hand, by replacing $k=0$ in Eq. (\ref{34.6}), we obtain the equation which describes  the behavior of the $h^{a}$-field
\begin{equation}
\dot{h}+H\left( h-C_{0}\right) =0,  \label{42}
\end{equation}%
whose solution, in terms of the redshift parameter, is given by 
\begin{equation}
h\left( z\right) =C_{0}+\left( h\left( 0\right) -C_{0}\right) \left(
1+z\right) .  \label{43}
\end{equation}%
Defining $\rho _{h}\left( z\right) =h\left( z\right) -C_{0}$, we can write $\rho _{h}\left( z\right) =\rho _{h}\left( 0\right) \left( 1+z\right) $ which
corresponds to a fluid with $\omega _{h}=-3/4$ and $q=-1/2$ (Dirac-Milne universe in Einstein-Hilbert 5D gravity, see later). Thus, the $h^{a}$-field behaves like a fluid that does not generate acceleration.

\section{EChS-$\mathfrak{B}_{5}$ and EChS-$\mathfrak{\tilde{B}}_{5}$ Cosmologies}

\subsection{EChS-$\mathfrak{B}_{5}$ Cosmology}

According to (\ref{35}), is straightforward to obtain
\begin{equation}
q=2\left( 1+\omega \right) \left( \frac{\rho }{\rho +3l^{2}H^{4}}\right) -1,
\label{44}
\end{equation}%
and, as we already saw, if $l=0$ we recover the Einstein scheme for which 
\begin{equation}
q=1+2\omega .  \label{45}
\end{equation}%
From (\ref{35}), the Hubble parameter turns out to be
\begin{equation}
H\left( z\right) =\frac{1}{l}\sqrt{\sqrt{1+\frac{l^{2}}{3}\rho \left(
z\right) }-1},  \label{46}
\end{equation}%
and using (\ref{39}), we write this parameter in the form
\begin{equation}
H\left( z\right) =\frac{1}{l}\sqrt{\sqrt{1+A\left( 1+z\right) ^{4\left(
1+\omega \right) }}-1},  \label{47}
\end{equation}%
where $A=l^{2}\rho \left( 0\right) /3$. The deceleration parameter, $q(z) =-1+\left( 1+z\right) dH/Hdz$, is given by the expression
\begin{equation}
q\left( z\right) =2\left( 1+\omega \right) \left( \frac{A\left( 1+z\right)
^{4\left( 1+\omega \right) }}{1+A\left( 1+z\right) ^{4\left( 1+\omega
\right) }-\sqrt{1+A\left( 1+z\right) ^{4\left( 1+\omega \right) }}}\right)
-1,  \label{48}
\end{equation}%
and it is direct to verify
\begin{eqnarray}
q\left( z\rightarrow \infty \right)  &\rightarrow &1+2\omega ,\text{ }
\label{49} \\
q\left( 0\right)  &=&2\left( 1+\omega \right) \left( \frac{A}{1+A-\sqrt{1+A}}%
\right) -1,\text{\ \ }  \label{50} \\
q\left( z\rightarrow -1\right)  &\rightarrow &3+4\omega ,  \label{51}
\end{eqnarray}
so that when $\omega >-1/2$ we find $q\left( z\right) >0$, i.e., a decelerated evolution. It is evident that when $\omega =-1$ we find $q(z) =-1$, i.e., a de Sitter evolution. An interesting detail is presented in (\ref{49}), this limit coincides with the behavior of $q$ when $l=0$. In this case, see (\ref{45}), $q=1+2\omega $, so that when $\omega =0$ one find $q=1$. Thus, for $\omega =0$, the limit given in (\ref{49}) tells us that early this fluid behaves like dark matter ($q=1$) and as stiff matter ($q=3$) at late times, see (\ref{51}). This situation, compared to its 4D equivalent, is unreasonable. The early behavior of the fluid is consistent with what is expected, but its late behavior is unrealistic. In
4D, stiff matter is a fluid that has very early relevance and a subsequent rapid ``dissolution" with cosmic evolution \textbf{\cite{Stiff}}.

From (\ref{47}), it is easy to verify that if $\omega >-1$ then $H(z\rightarrow \infty) \rightarrow \infty$ and $H\left( z\rightarrow-1\right) \rightarrow 0$. And if $\omega =-1$, $H\left(z\rightarrow-1\right) \rightarrow \left( 1/l\right) \sqrt{\sqrt{1+A}-1}$. Additionally, according to (\ref{45}), $\omega=0$ leads to $q=1$ (in 4D, $\omega=0$ leads to $q=1/2$), $\omega =-1$ leads to $q=-1$, $\omega <-1$ leads to $q<-1$ (phantom evolution). If $\omega =-1/2$ one find  $q=0$, we have the Dirac-Milne universe. But, if we look (\ref{48})
\begin{equation}
\omega =-\frac{1}{2}\ \ \Longrightarrow\ \ q\left( z\right) =\frac{A\left(
1+z\right) ^{2}}{1+A\left( 1+z\right) ^{2}-\sqrt{1+A\left( 1+z\right) ^{2}}}%
-1,  \label{52}
\end{equation}
we have an evolution with non-zero acceleration.

\begin{figure}[h]
\centering
\includegraphics[scale=1]{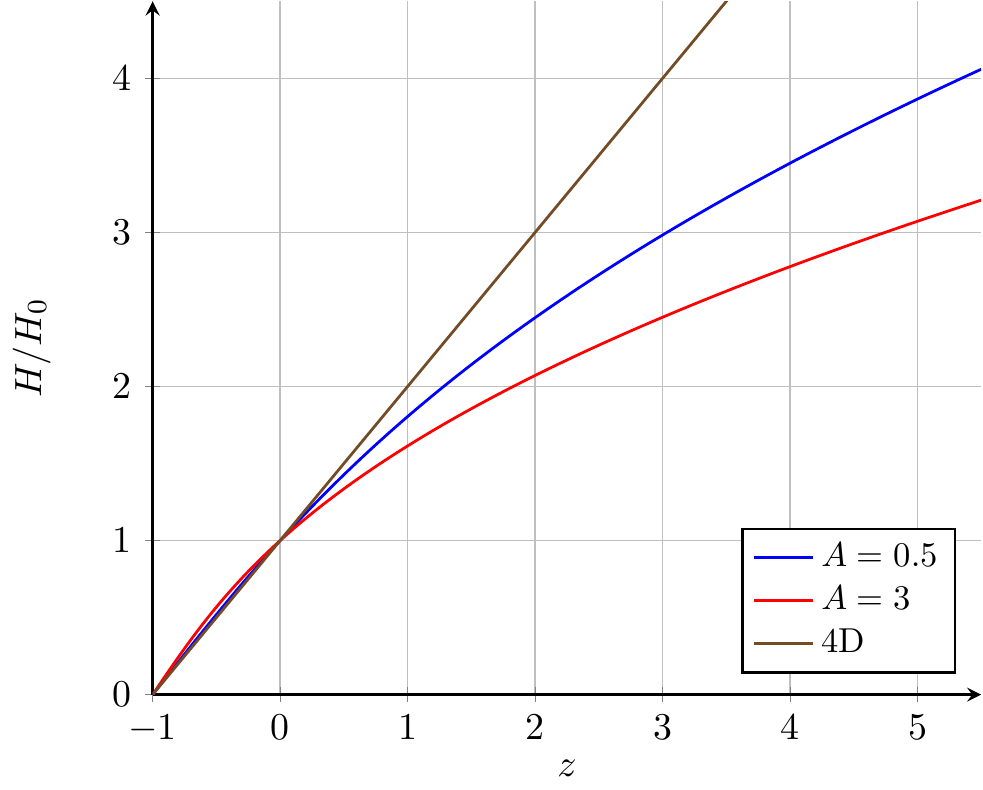}
\caption{The Hubble parameter for $\protect\omega =-1/2$. Here, $A=0.5$ (\textit{blue line}) and $A=3$ (\textit{red line}). The 4D-case is the 
\textit{black line}.}
\label{fig01}
\end{figure}

\begin{figure}[h]
\centering
\includegraphics[scale=1]{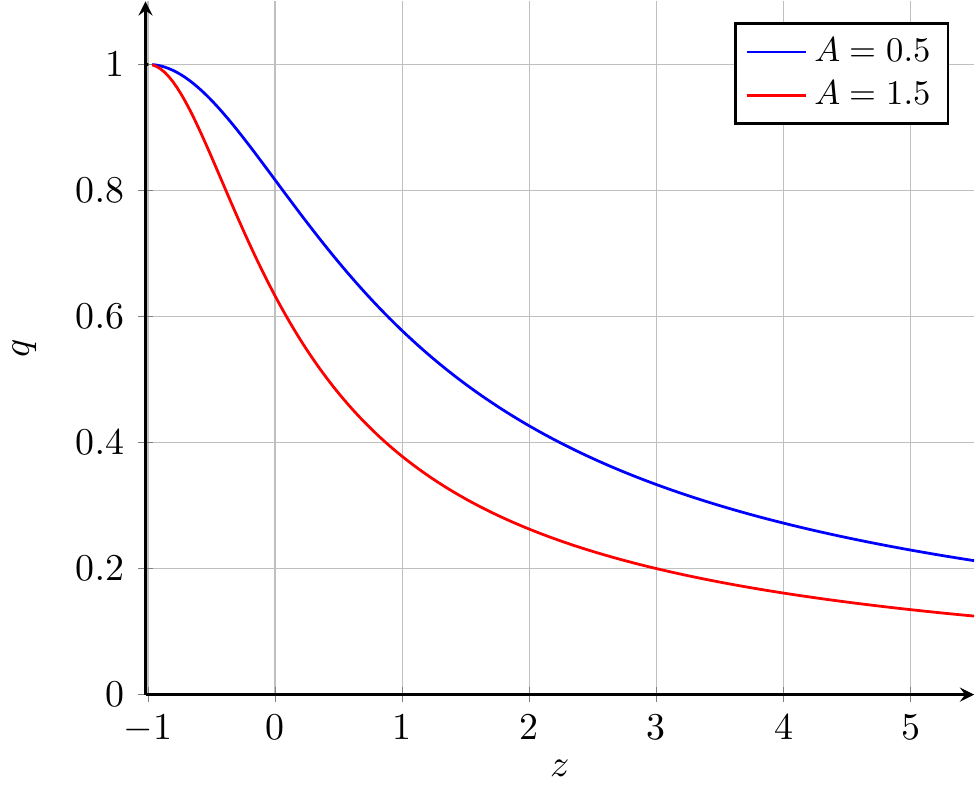}
\caption{The deceleration parameter for $\protect\omega =-1/2$. Note that $q=0 $ is recovered when $z\rightarrow \infty$. Here, $A=0.5$
(\textit{blue line}) and $A=1.5$ (\textit{red line}).}
\label{fig02}
\end{figure}

We rescue two results from this Section: for $\omega =0$ we obtain an early evolution as we expect, that is, an evolution driven by dark matter. However, the late behavior it's like stiff matter and this, as we said, is not realistic. The other result is an accelerated Dirac-Milne universe.

So, this is the Cosmology that seems relevant from EChS-$\mathfrak{B}_{5}$ gravity.

\subsection{EChS-$\mathfrak{\tilde{B}}_{5}$ Cosmology}

Writing the left equation of (\ref{36'}) in the form
\begin{equation}
\left( \frac{2}{l^{2}}-H^{2}\right) H^{2}=\frac{1}{3l^{2}}\rho ,\label{53}
\end{equation}
we immediately remember the DGP model (4D), Refs. \cite{DGP} and \cite{Lepedev}. In the flat case ($k=0$) and $8\pi G=c=1$, this model is
\begin{equation}
\left( H-\frac{\epsilon }{r_{c}}\right) H=\rho ,  \label{54}
\end{equation}
where $\epsilon =\pm 1$ and $r_{c}$ is a characteristic scale. In the absence of $\rho $, we have a self-accelerating solution for $\epsilon =1$,
that is, $H=1/r_{c}$. In the present case
\begin{equation}
\rho =0\ \text{ leads to }\ H=\frac{\sqrt{2}}{l},  \label{55}
\end{equation}
an analogous situation to that of said model.

The solution for the Hubble parameter is given by
\begin{equation}
H_{\pm }\left( z\right) =\frac{1}{l}\sqrt{1\pm \sqrt{1-\frac{l^{2}}{3}\rho
\left( z\right) }},  \label{56}
\end{equation}%
i.e.,
\begin{equation}
H_{\pm }\left( z\right) =\frac{1}{l}\sqrt{1\pm \sqrt{1-\left( \frac{1+z}{1+%
\bar{z}}\right) ^{4\left( 1+\omega \right) }}},  \label{57}
\end{equation}%
where
\begin{equation}
1+\bar{z}=\frac{1}{A^{1/4\left( 1+\omega \right) }},  \label{58}
\end{equation}
recalling that $A=l^{2}\rho \left( 0\right) /3$. According to (\ref{57}), we note that if $\omega>-1$, then
\begin{equation}
z\leq \bar{z}\text{ \ \ and \ \ }H_{\pm }\left( \bar{z}\right) =\frac{%
1}{l},  \label{59}
\end{equation}
and given that
\begin{eqnarray}
\text{if }A &>&1\text{ we have }-1<\text{\ }\bar{z}<0,  \label{60} \\
\text{if \ }A &<&1\text{ we have  }\bar{z}>0,  \label{61}
\end{eqnarray}
then, if $\bar{z}$ belongs to the past, we have an cosmological evolution from $\bar{z}$, which implies, $H_{\pm }\left( \bar{z}\right) =1/l$ to $z\rightarrow -1$, which implies $H_{\pm}\left( -1\right)= \sqrt{2}/l$. On the contrary, if $\bar{z}$ is in the future, we have a late evolution from $-1<\bar{z}<0$, which implies, $H_{\pm }\left( \bar{z}\right)=1/l$ to $z\rightarrow -1$, which implies, $H_{\pm }\left(-1\right) =\sqrt{2}/l$.

As shown in the flat case ($k=0$), the pressure associated with this density, $\rho ^{\left( h\right) }=3l^{2}H^{4}$, satisfies the equations $p^{\left( h\right) }=q\rho ^{\left( h\right) }$. We can see an analogous situation in 4D, that is, two non-interacting fluids, a dark matter $\rho$ fluid plus a holographic fluid $\rho_\text{hol}$ like $\rho _\text{hol}\sim H^{2}$ \cite{Miao Li}.
\begin{equation}
3H^{2}=\rho +\rho_\text{hol}\text{ \ \ },\text{ \ \ }\rho_\text{hol}=3c^{2}H^{2}\ \rightarrow\  \dot{\rho}_\text{hol}=-2H\left( 1+q\right)\rho_\text{hol},
\end{equation}
and
\begin{equation}
\dot{\rho}+3H\left( 1+\omega \right) \rho =0\text{ \ \ },\text{ \ \ }\dot{\rho}_\text{hol}+3H\left( 1+\omega _{h}\right) \rho _\text{hol}=0.
\end{equation}%
Then, it is straightforward to show
\begin{equation}
\omega_\text{hol}=\frac{2}{3}\left( q-\frac{1}{2}\right) ,
\end{equation}
so that for  $q\gtrless 1/2$ we have $\omega_\text{hol}\gtrless 0$. In the present situation, 5D, $q\gtrless 0$ leads to $\omega_{h}\gtrless 0$. In good accounts, there is a clear option to have dark matter or dark energy
from the EChS-$\mathfrak{\tilde{B}}_{5}$ framework. A question. Is this holographic analogy viable? So, this is the Cosmology that seems relevant from EChS-$\mathfrak{\tilde{B}}_{5}$.

\section{Concluding Remarks}

In this paper we have considered EChS gravity instead of GR to describe the expansion of a flat 5-dimensional universe. Although some solutions of this kind have been found in Refs.\cite{gomez} and \cite{gomez1}, in this article a wider cosmological analysis is performed and, in addition, the connection between the EChS theory and the schemes for EChS-$\mathfrak{\tilde{B}}_{5}$ and EChS-$\mathfrak{B}_{5}$ gravities is explicitly shown. This is achieved by a correct choice of constants in the theory, which is guided by certain assumptions on the algebra describing the symmetry of the system.

The gravitational theory of EChS includes, in besides the usual geometrical fields, two fields $k^{ab}$ and $h^{a}$. The $k^{ab}$-field was assumed null by virtue of the gauge freedom and the $h^{a}$ field was presented as a perfect fluid. It is important to note that the identification of the field $h^{a}$as a fluid is not entirely accurate, since from (\ref{34-4}) it is straightforward to see that the quantity $\rho ^{(h)}$ takes a negative value in the EChS-$\mathfrak{B}_{5}$ framework. On the other hand, in the EChS-$\mathfrak{\tilde{B}}_{5}$ case, with $\rho ^{(h)}>0$ the interpretation of $h^{a}$ as a perfect fluid is appropriate.

We have discussed the $\pm 3l^{2}H^{4}$ term in both EChS-$\mathfrak{B}_{5}$ ($``+"$) and EChS-$\mathfrak{\tilde{B}}_{5}$ ($``-"$) schemes and we have showed the resultant cosmologies. In the EChS-$\mathfrak{B}_{5}$ case, we highlight an accelerating Dirac-Milne universe and a fluid, for which $\omega =0$, that early behaves like dark matter but later behaves like stiff matter. In the EChS-$\mathfrak{\tilde{B}}_{5}$ case we highlight two
schemes, a self-accelerating scheme (an analogue to the DGP model in 4D) beside a holographic analogy. This last interpretation is an open issue.

Finally, a compactification 5D to 4D would be interesting to do. In particular, we wonder about the consequences of the $h$-field in the resulting cosmology in 4D. What would we expect from this compactification? This answer is not obvious. In the present work, 5D, we show that the $h$-field behaves like a fluid that does not generate acceleration (Dirac-Milne universe), see Eq. (\ref{43}) and its interpretation as a energy density $\rho _{h}\left( z\right) \sim (1+z)$. However, in a 4D
context, it is possible to conjecture the existence of accelerated solutions, eventually driven by the $h$-field. For instance, in the framework of braneworld cosmology, in Refs. \cite{Brane01,Brane} the $h$-field was associated with a scalar field, exhibiting the behavior of cosmological constant (dark energy). This background could then lead to a response to the concern presented. Of course, the behavior of the $h$-field in 4D, and its possible physical interpretations, will depend on the mechanism of dimensional reduction under consideration.

\textbf{Acknowledgements}

This work was supported in part by Grant \# R12/18  from Direcci\'{o}n de Investigaci\'{o}n, Universidad de Los Lagos (FG), in part by Direcci\'{o}n de Investigaci\'{o}n, Universidad San Sebasti\'{a}n (CQ), and  in part by FONDECYT through Grant No. 1180681 from the Government of Chile (PS).

\section{Appendix: Derivation of the EChS Field\break Equa\-tions for the 5D-FLRW metric}

In this Appendix, the derivation of the EChS field equations for the 5D-FLRW metric is presented.

In five dimensions, the FLRW metric is given by (see e.g., Ref. \cite{Cataldo})
\begin{equation}
\mathrm ds^{2}=-\mathrm dt^{2}+a^{2}(t)\left[ \frac{\mathrm dr^{2}}{1-kr^{2}}+r^{2}\left(\mathrm d\theta
_{1}^{2}+\sin^{2}\theta _{1}\,\mathrm d\theta _{2}^{2}+\sin^{2}\theta
_{1}\sin^{2}\theta _{2}\,\mathrm d\theta _{3}^{2}\right) \right] ,  \label{a1}
\end{equation}
where $a(t)$\ is the cosmic scale factor and $k=+1,0,-1$ describes spherical, flat and hyperbolic spaces, respectively. This line element can
be written in terms of the vielbein $e^{a}$ as
\begin{equation}
\mathrm ds^{2}=\eta _{ab}e^{a}e^{b},  \label{a2}
\end{equation}
where $\eta _{ab}$ is the Minkowski metric $\eta _{ab}=\mathrm{diag}\left(
-1,+1,+1,+1,+1\right) $ and 
\begin{equation*}
e^{0}=\mathrm dt\text{ \ \ },\text{ \ }e^{1}=\frac{a(t)}{\sqrt{1-kr^{2}}}\,\mathrm dr\text{ \
\ \ },\text{ \ }e^{2}=a(t)r\,\mathrm d\theta _{1},
\end{equation*}%
\begin{equation}
e^{3}=a(t)r\sin\theta _{1}\,\mathrm d\theta _{2}\text{ \ \ \ },\text{ \ \ }%
e^{4}=a(t)r\sin\theta _{1}\sin\theta _{2}\,\mathrm d\theta _{3}.  \label{a3}
\end{equation}
If the null torsion condition $T^{a}=0$ is imposed, the spin connection $\omega ^{ab}$ can be obtained from Cartan's first structural equation $T^{a}=\mathrm de^{a}+\omega _{\ b}^{a}e^{b}$. In fact, the non-null spin components are given by
\begin{equation}
\omega _{\text{ }p}^{0}=\omega _{\text{ }0}^{p}=He^{p};\text{ \ \ \ }%
p=1,2,3,4;
\end{equation}
\begin{equation}
\omega _{\text{ }p}^{1}=-\omega _{\text{ }1}^{p}=\frac{\sqrt{1-kr^{2}}}{ar}\,
e^{p};\text{ \ \ \ }p=2,3,4;
\end{equation}
\begin{equation}
\omega _{\text{ }p}^{2}=-\omega _{\text{ }2}^{p}=-\frac{1}{ar}\cot \theta
_{2}\,e^{p};\text{ \ \ \ }p=3,4;
\end{equation}%
\begin{equation}
\omega _{\text{ }4}^{3}=-\omega _{\text{ }3}^{4}=-\frac{1}{ar}\frac{1}{\sin
\theta _{1}}\cot \theta _{2}\,e^{4},
\end{equation}%
where $H=\dot{a}/a$ is the Hubble parameter. By replacing these quantities in Cartan's second structural equation $R_{\text{ }b}^{a}=\mathrm d\omega _{\ b}^{a}+\omega _{\ c}^{a}\omega_{\ b}^{c}$, the 2-form curvature
is obtained
\begin{equation}
R^{0p}=\left( \dot{H}+H^{2}\right) e^{0}e^{p};\text{ \ \ \ }p=1,2,3,4;
\label{a5}
\end{equation}%
\begin{equation}
R^{pq}=\left( H^{2}+\frac{k}{a^{2}}\right) e^{q}e^{p};\text{ \ \ }%
p,q=1,2,3,4.  \label{a6}
\end{equation}

On the other hand, the field $h^{a}$, also present in the EChS field equations, should be modeled according to the spatial-temporal symmetries of the FLRW metric; i.e., the 2-rank tensor $h_{\mu \nu }$ in $h^{a}=h_{\ \mu}^{a}e^{\mu }=\eta ^{a\mu }h_{\mu \nu }e^{\nu }$ should be shaped according to the cosmological principle, respecting spatial homogeneity and isotropy. Thus, following Refs. \cite{gomez}, \cite{gomez1}, \cite{weinberg} we write down
\begin{equation}
h^{0}=C_{0}e^{0},
\end{equation}%
\begin{equation}
h^{p}=h(t)e^{p}\text{ \ },\text{ \ }p=1,2,3,4;
\end{equation}%
where $C_{0}$ is a constant and $h(t)$ is a time-dependent function. It is useful to calculate the exterior covariant derivative of this field $\mathrm{D}h^{a}$, which results in
\begin{equation}
\mathrm{D}h^{0}=0,
\end{equation}%
\begin{equation}
\mathrm{D}h^{p}=\left[ \dot{h}+H\left( h-C_{0}\right) \right] e^{0}e^{p};%
\text{ \ \ }p,q=1,2,3,4.
\end{equation}

In addition, we assume that the material content of the model is described by perfect fluids, associated with the tensors
\begin{equation}
T_{\mu }^{\ \nu }=\mathrm{diag}(-\rho ,p,p,p,p),
\end{equation}%
\begin{equation}
T_{\mu }^{(h)\nu }=\mathrm{diag}(-\rho
^{(h)},p^{(h)},p^{(h)},p^{(h)},p^{(h)}),  \label{a10}
\end{equation}%
where $\rho (t)$ and $p(t)$ are the usual energy density and pressure, and, $\rho ^{(h)}(t)$ and $p^{(h)}(t)$ represent their counterparts corresponding to the $h^{a}$-field.

The EChS field equations describing the 5D cosmological dynamics are obtained by substituting (\ref{a5})- (\ref{a10}) in equations (\ref{17})-(\ref{19}) 
\begin{equation}
6\left( H^{2}+\frac{k}{a^{2}}\right) \pm 3l^{2}\left( H^{2}+\frac{k}{a^{2}}%
\right) ^{2}=\kappa \rho ,  \label{a11}
\end{equation}%
\begin{equation}
3\left( \dot{H}+2H^{2}+\frac{k}{a^{2}}\right) \pm 3l^{2}\left( \dot{H}%
+H^{2}\right) \left( H^{2}+\frac{k}{a^{2}}\right) =-\kappa p,  \label{a12}
\end{equation}
\begin{equation}
\rho ^{(h)}=\mp \frac{3l^{2}}{\kappa }\left( H^{2}+\frac{k}{a^{2}}\right)
^{2},
\end{equation}%
\begin{equation}
p^{(h)}=\pm \frac{3l^{2}}{\kappa }\left( \dot{H}+H^{2}\right) \left( H^{2}+%
\frac{k}{a^{2}}\right),
\end{equation}%
\begin{equation}
\left( H^{2}+\frac{k}{a^{2}}\right) \left[ \dot{h}+H\left( h-C_{0}\right) %
\right] =0.  \label{a13}
\end{equation}

\end{document}